# Cost-Control in Display Advertising: Theory vs Practice

Anoop R Katti
anoop.katti@zalando.de
Zalando SE
Berlin, Germany

Rui C. Gonçalves
rui.goncalves@zalando.de
Zalando SE
Berlin, Germany

Rinchin Iakovlev
rinchin.iakovlev@zalando.de
Zalando SE
Berlin, Germany

## ABSTRACT

In display advertising, advertisers want to achieve a marketing objective with constraints on budget and cost-per-outcome. This is usually formulated as an optimization problem that maximizes the total utility under constraints. The optimization is carried out in an online fashion in the dual space – for an incoming Ad auction, a bid is placed using an optimal bidding formula, assuming optimal values for the dual variables; based on the outcome of the previous auctions, the dual variables are updated in an online fashion. While this approach is theoretically sound, in practice, the dual variables are not optimal from the beginning, but rather converge over time. Specifically, for the cost-constraint, the convergence is asymptotic. As a result, we find that cost-control is ineffective. In this work, we analyse the shortcomings of the optimal bidding formula and propose a modification that deviates from the theoretical derivation. We simulate various practical scenarios and study the cost-control behaviors of the two algorithms. Through a large-scale evaluation on the real-word data, we show that the proposed modification reduces the cost violations by 50%, thereby achieving a better cost-control than the theoretical bidding formula.

## 1 INTRODUCTION

In online display advertising, advertisers run Ad campaigns with a goal of achieving some marketing objectives, e.g. increasing the awareness/engagement of their brand, boosting sales, etc. Online Ads are often sold using auction mechanisms where the competing Ads place a bid and are charged for a user view or a click.

Advertisers have a certain advertising budget that they wish to spend in a specific period of time and they wish to control the cost-per-outcome (e.g. cost-per-view, cost-per-click). While the budget represents the total volume of Ad spend, cost-per-outcome represents the efficiency with which the outcome is achieved. In fact, it is not uncommon for advertisers to set up campaigns with high advertising budgets and steer them on cost-per-outcome. This makes it important to meet both the budget and the cost constraints.

In this paper, we consider a scenario where advertisers run a marketing campaign for a fixed period of time to achieve a marketing objective while keeping the total advertising spend within a specified budget and the cost-per-view under a specified target-cost. Such a scenario has been widely studied in the literature of Real-time Bidding [9, 11, 16]. A common way to approach it is to formulate it as an optimization problem that maximizes the advertiser utility (cost-adjusted value) under constraints. Each incoming Ad opportunity is supposed to have an expected value to the advertiser. The constraints capture the upper bound on the total Ad spend and the cost-per-outcome. Note that when competing for the current Ad opportunity, we do not have access to the future Ad opportunities. Consequently, based on the historical outcomes, the optimization is carried out in an online fashion and in the dual space [3]. Assuming the current values of the dual variables (computed from the online optimization) are optimal, an optimal bidding formula is derived that maximizes the advertiser utility for each Ad opportunity, accounting for the current violation of constraints. Based on the outcomes of the historical auctions, the dual variables associated with the constraints are updated towards their optimal values in an online fashion.

This approach is theoretically sound. However, the values of the dual variables are not optimal from the beginning. In fact, they converge to the optimal value over time. In particular, the dual variable associated with the cost-constraint converges asymptotically. As a result, we find that in practice, cost-control is ineffective. Furthermore, say the dual variable associated with cost-control has (almost) converged to the optimal value. If the market conditions change or if the campaign settings are modified by the advertisers, the algorithm would take a very long time to stabilise again to the new optimal value. In the rest of the paper, we analyse the optimal bidding formula and highlight its shortcomings when applied in an online setting. Based on the analysis, we propose a change to the bidding formula that deviates from the theoretical derivation. We simulate several practical scenarios and compare the cost-control behaviors of the two algorithms. Lastly, through a large-scale evaluation on the real-world data, we demonstrate that the proposed modification violates cost constraints on 50% fewer campaigns as compared to the theoretical bidding formula.

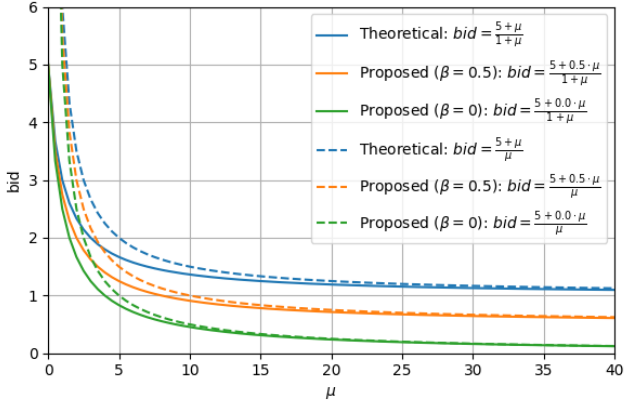

Figure 1: Bid variation with $\mu$ – Theoretical (blue), Proposed (orange, green); solid and dashed lines indicate $\alpha = 1$ and $0$.

## 2 RELATED WORK

Real-time Bidding has been extensively studied in the scientific literature [5, 17, 18]. Several of the existing works in real-time bidding address the topic of bidding in repeated auctions with budget constraints [1, 2, 6, 8, 10, 14]. The approaches can be grouped in three categories: (i) solving an online optimization using dual online mirror descent [3, 9], (ii) using a Proportional–Integral–Derivative (PID) controller to control the spend-rate around some target [15–17], and (iii) maximizing the total reward using reinforcement-learning in either the primal [5] or the dual space [11]. In fact, in [4], the authors prove that dual-based PI controllers are the same as solving the dual optimization using online mirror descent with momentum.

We consider not just the budget constraints, but also the cost-per-outcome constraints. Two studies that are close to the problem we consider are from LinkedIn [9] and Alibaba [16]. Similar to the works on budget constraints, they formulate and solve an optimization problem to derive the optimal bidding formula that captures both the budget and the cost constraints and optimize the dual variables using either the online mirror descent [9] or a PID controller [16]. We base our work primarily on these two studies.

The bidding formula is optimal only if the dual variables have converged to their optimal values. However, in practice, the dual variables are not optimal from the beginning. Specifically, the dual variable associated with cost-control converges only asymptotically. In this work, we propose a modification to the bidding formula that deviates from the theoretical derivation and show that in practice, it is more effective than the optimal bidding formula. To the best of our knowledge, there is no previous work that studies this topic.

Other relevant literature in the field include building auction simulators [12, 13] and handling more extensive constraints such as achieving a certain distribution of users [7].

## 3 METHOD

We consider the problem of maximizing the total advertiser utility (cost-adjusted value) given a target budget and a target cost-per-view. Ad opportunities are sold using an auction mechanism. Similar to the literature [9, 16], we derive the optimal bidding formula for participating in Ad auctions. We then share a practical issue with cost-control and propose a modification that deviates from the theoretical derivation, but works in practice.

### 3.1 Optimal Bidding

Consider a single Ad campaign with an advertising budget of $B$ and a target cost-per-view of $C_{\text{view}}$. Let the campaign participate in $N$ Ad opportunities (i.e. Ad auctions). For the $i^{th}$ Ad opportunity, let $\text{bid}_i$ be the bid placed by this Ad campaign in the auction, $x_i^{\text{bid}_i}$ be the resulting probability of a user view, $v_i$ be the value of the view to the advertiser and $c_i$ denote the cost of the view. We define the optimization objective as $\sum_{i=1}^{N} x_i^{\text{bid}_i} \cdot (v_i - \alpha \cdot c_i)$, where $0 \leq \alpha \leq 1$ is the fraction of cost considered in the objective. If $\alpha = 1$, we optimize for utility and if $\alpha = 0$, we optimize for value. The optimization problem is formulated as:

$$\max_{x^{\text{bid}} \in [0,1]^N} \sum_{i=1}^{N} x_i^{\text{bid}_i} \cdot (v_i - \alpha \cdot c_i) \quad (1a)$$

$$\text{subject to} \quad \sum_{i=1}^{N} x_i^{\text{bid}_i} \cdot c_i \leq B, \quad (1b)$$

$$\sum_{i=1}^{N} x_i^{\text{bid}_i} \cdot c_i \leq C_{\text{view}} \sum_{i=1}^{N} x_i^{\text{bid}_i}. \quad (1c)$$

For second-price auctions, solving the above optimization problem results in the following optimal bidding formula [9, 16]:

$$\text{bid}_i = \frac{v_i + \mu \cdot C_{\text{view}}}{\alpha + \lambda + \mu}, \quad (2)$$

where $\lambda > 0$ and $\mu > 0$ are dual variables associated with the budget and the cost-per-view constraints respectively.

Eq 2 assumes that the dual variables are at their optimal values, $\lambda^*$ and $\mu^*$. However, $\lambda^*$ and $\mu^*$ are not known from the beginning. Rather, the dual variables are updated towards their optimal values in an online-fashion. Concretely, after each mini-batch of $n$ auctions, $i^t : i^{t+1}$, the $x_i$'s materialise into binary outcomes, i.e. viewed or not viewed. Let $\epsilon_\lambda$ and $\epsilon_\mu$ be the learning rates for the online updates. Then $\lambda$ and $\mu$ are updated as [1]:

$$\lambda^{t+1} = \max\left(\lambda^t + \frac{\epsilon_\lambda}{n} \sum_{i^t}^{i^{t+1}} \left(x_i \cdot c_i - \frac{B}{N}\right), 0\right) \quad (3a)$$

$$\mu^{t+1} = \max\left(\mu^t + \frac{\epsilon_\mu}{\sum_{i^t}^{i^{t+1}} x_i} \sum_{i^t}^{i^{t+1}} x_i \left(c_i - C_{\text{view}}\right), 0\right) \quad (3b)$$

### 3.2 Practical Issue with Cost-control

In this section, we analyse the optimal bidding formula. Suppose budget is not an active constraint, i.e. $B$ is sufficiently high. In this case, $\lambda^t = \lambda^* = 0$. Say, advertisers want a return on Ad-spend of $R > 1$. Then, the expected value, $E[v] = R \cdot C_{\text{view}}$ and the (average)

---

[1] For spend-control, we use the Model Predictive Control version as in [9]. For cost-control, we use momentum in order to be robust against outliers; this is particularly important when the overall cost-per-view is lower than the target, i.e. the constraint is inactive and a single outlier can activate the constraint and reduce the bid.

bid = $C_{\text{view}} \cdot \frac{R+\mu}{\alpha+\mu}$. Fig 1 shows how bid varies with $\mu$ for $R = 5$, $C_{\text{view}} = 1$ and $\alpha = 1$ (solid blue). We can see that bid $\geq C_{\text{view}}$. In fact, the equality, bid* = $C_{\text{view}}$, happens at the optimal value, $\mu^* = \infty$. With $\alpha = 0$ (dashed blue in Fig 1), the lower-bound on bid still holds.

In an online optimization setting, the algorithm starts with a $\mu < \infty$ and reaches the optimal value asymptotically over a long period of time. As discussed in Sec 4.1, due to several factors that are outside the control of the algorithm, the cost-per-view may exceed the specified target, $C_{\text{view}}$. If the bid does not drop below $C_{\text{view}}$, then in such cases, it may be infeasible to meet the cost constraint. Consequently, the bidding algorithm does not achieve an effective cost-control.

### 3.3 Improving Cost-control

As before, suppose budget is sufficiently high and is, therefore, not an active constraint. Intuitively, if the current cost-per-view is greater than $C_{\text{view}}$, $\mu$ should increase over some $\mu' < \infty$, bid should drop below $C_{\text{view}}$ and the cost-per-view should gradually decrease to $C_{\text{view}}$. Further, once the cost is under control, i.e. cost-per-view equals $C_{\text{view}}$, $\mu$ should stabilise at $\mu'$. Let's consider introducing an artificially discounted $C'_{\text{view}} = \beta \cdot C_{\text{view}}, \beta \in [0, 1)$ in the original optimization (Eq 1c). With this change, the bid can now go below $C_{\text{view}}$ and this happens at some $\mu' < \infty$. However, even after the cost-per-view decreases to $C_{\text{view}}$, the (average) error in Eq 3b continues to remain positive (as the errors are computed against $C'_{\text{view}}$). Consequently, $\mu$ will not stabilise at $\mu'$, but instead keeps increasing (asymptotically) to $\infty$. This unnecessarily pushes the cost-per-view below the real target, $C_{\text{view}}$ and negatively impacts the value generated to the advertiser.

Based on this intuition, we propose the following change that does not conform to the theory, but takes the best of both worlds: modify the bidding formula as if there is a discounted target cost-per-view and leave the online updates in Eq 3a and Eq 3b unchanged. This results in the alternative bidding formula shown below:

$$\text{bid}_i = \frac{v_i + \mu \cdot \beta \cdot C_{\text{view}}}{\alpha + \lambda + \mu} \quad (4)$$

Fig 1 visualises how bid varies with $\mu$ for $R = 5$, $C_{\text{view}} = 1$, $\alpha = 1$ and for two different values of $\beta$ (solid orange: $\beta = 0.5$, solid green: $\beta = 0$). We can see that the (average) bid = $C_{\text{view}} \cdot \frac{R+\mu \cdot \beta}{\alpha+\mu} \geq \beta \cdot C_{\text{view}}$. This lower-bound also holds for $\alpha = 0$ (dashed orange and dashed green in Fig 1). This allows the bid to go below $C_{\text{view}}$, thereby, bringing down the cost when necessary. Once the cost reaches the target, $C_{\text{view}}$, the (average) error in Eq 3b goes to zero, $\mu$ stabilises and the cost remains at the target. While this modification deviates from the theory, we argue that in an online optimization setting, it enables the bidding algorithm to achieve effective cost-control.

## 4 EVALUATION

In this section, we will first analyse the cost-control behaviors of the two bidding algorithms in different practical scenarios – the theoretically derived bidding formula and the proposed bidding formula. Then, we will perform quantitative comparison on large-scale real-world Ad opportunities and campaigns.

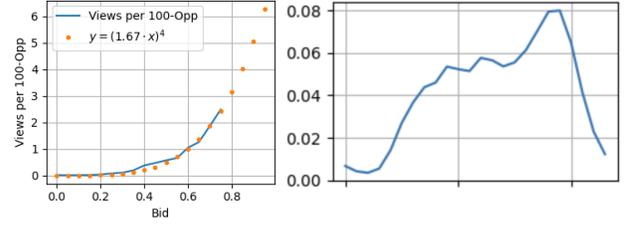

(a) View-rate vs Bid    (b) Simulation Traffic distribution

### 4.1 Simulations on Synthetic Data

The goal of the simulations is to qualitatively study the cost-control behaviors of the algorithms in different scenarios that are relevant in practice.

*4.1.1 Set-up:* We consider a single campaign with $B$ as the budget for one day and $C_{\text{view}}$ as the target cost-per-view. For simplicity of simulation, we choose the awareness objective, i.e. obtain user views; however, the analysis on cost-control should generalise to other marketing objectives such as engagement (i.e. clicks) and purchase (i.e. sales). In our simulations, we maximise utility, i.e. $\alpha = 1$. The value of a view is set to, $v_i = R \cdot C_{\text{view}}$, where $R > 1$ is a constant representing the advertisers' return on Ad spend. From real auctions, we have constructed a relationship between bid and view-rate. This is a property of the market. This relationship is visualised in Fig 2a. Notice that as bid increases, view-rate increases (polynomial with a degree 4). The price per view is 95% of the bid to emulate a second-price auction. We consider $O(10^6)$ Ad opportunities and that they are distributed as shown in Fig 2b to mimic the real traffic distribution. The dual variables, $\lambda$ and $\mu$, are initialised such that the initial bid is $C_{\text{view}}$ (and $\lambda = \mu$). Further, they are updated every 1 minute with learning rate recommended in [9].

*4.1.2 Variants:* We consider three variants
  (i) *Max-cap*: here, we apply a hard cap on the bid, where the cap is the target cost-per-view; specifically, $\text{bid}_i = \min(v_i/(1 + \lambda), C_{\text{view}})$.
  (ii) *Cost-control-theoretical*: here, we employ the bidding formula from Eq 2 to achieve the target cost-per-view.
  (iii) *Cost-control-practical*: here, we employ the proposed bidding formula from Eq 4 to achieve the target cost-per-view. We set $\beta = 0$ as this provides the most flexibility to the algorithm to drop the bid all the way to 0, if necessary.

*4.1.3 Metrics:* We visualize the evolution of the dual variables, $\lambda$ and $\mu$, the actual and target cost-per-view and the actual and target spend.

*4.1.4 Practical Scenario 1: Budget is the active constraint.* Consider a scenario where budget is the active constraint, i.e. target cost-per-view is set to some high value. In this case, the optimal values for the dual variables are: $(R-1) < \lambda^* < \infty$ for all variants and $\mu^* = 0$ for variants (ii) and (iii) (variant-(i) does not have $\mu$). We find that after a few iterations, all three variants converge to the optimal values. This is visualised in Fig 3.

*4.1.5 Practical Scenario 2: Cost is the active constraint.* Consider a scenario where cost-per-view is the active constraint, i.e. budget is sufficiently high. In this case, the optimal values for the dual variables are $\lambda^* = 0$ for all variants; variant-(i) does not have $\mu$; for variant-(ii), $\mu^* = \infty$; for variant-(iii), $\mu^* = R - 1$.

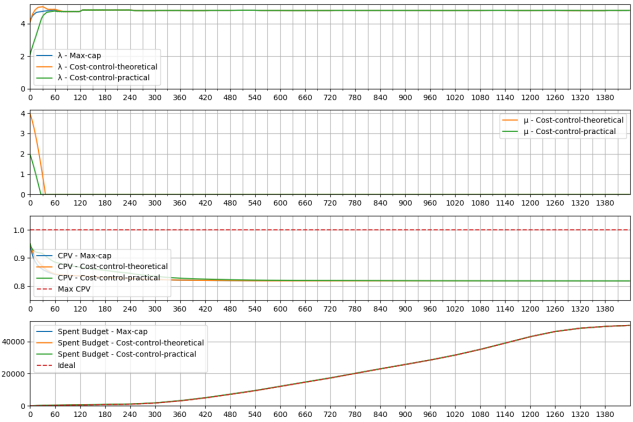

Figure 3: Scenario 1 – Budget is the active constraint

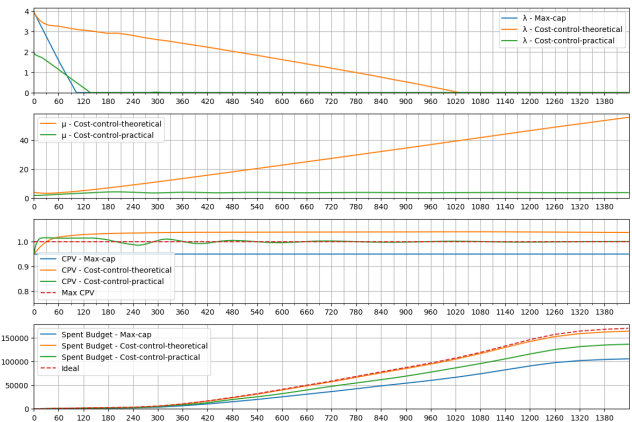

Figure 4: Scenario 2 – Cost is the active constraint

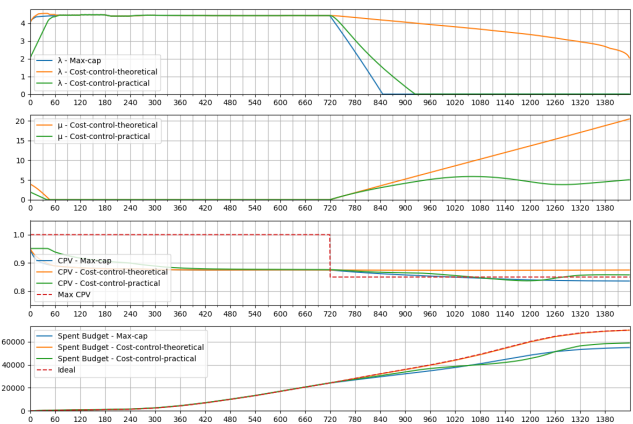

Figure 5: Scenario 3 – Budget is the active constraint and a cost-sensitive advertiser decreases the target cost-per-view

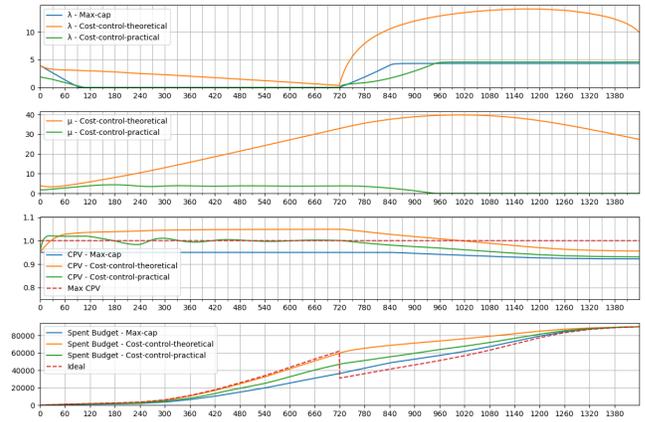

Figure 6: Scenario 4 – Cost is the active constraint and advertiser moves the budget to a different channel

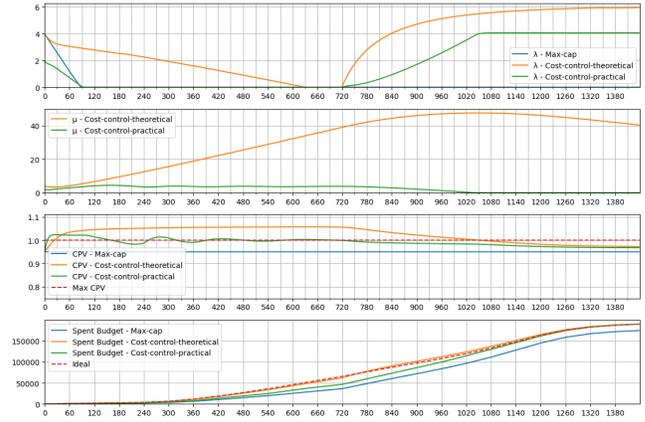

Figure 7: Scenario 5 – Cost is the active constraint and the traffic becomes 2x

We find that variants-(i) and (iii) converge to their optimal values after a few iterations; variant-(ii) converges to $\lambda^t = 0$, but, as expected, $\mu^t$ keeps growing. More importantly, in variant-(ii), since $\mu^t$ never reaches to its optimal value, the cost-per-view is always higher than the target. On the other hand, *Max-cap* and *Cost-control-practical* reach the target cost-per-view. This is shown in Fig 4.

*4.1.6 Practical Scenario 3: Budget is the active constraint and a cost-sensitive advertiser decreases the target cost-per-view.* Consider a scenario where the advertiser is highly sensitive to the costs (and ultimately, return on Ad spend). At the beginning of the campaign, budget is the active constraint. However, the advertiser realises during the campaign that they are not achieving the desired return on Ad spend and decreases the target cost-per-view. Now the cost becomes the active constraint and $\mu^t$ should go from $\mu^* = 0$ to $\mu^* = \infty$ in the case of variant-(ii). This takes a long time and it leads

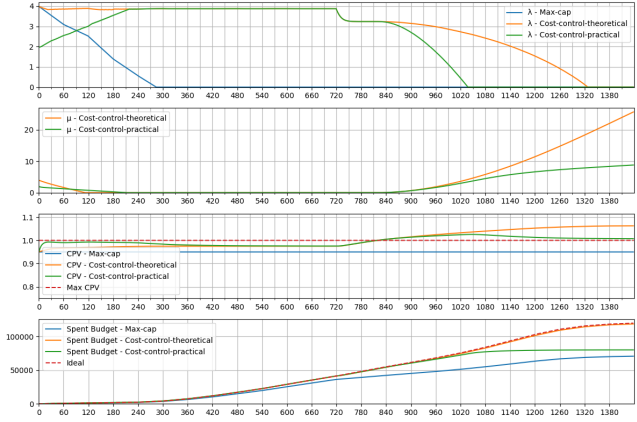

Figure 8: Scenario 6 – Budget is the active constraint and the traffic drops to 0.5x

to a higher violation of the cost-per-view constraint as compared to other variants. This is visualised in Fig 5.

*4.1.7 Practical Scenario 4: Cost is the active constraint and advertiser moves the budget to a different channel.* Consider a scenario where cost is the active constraint and that *Cost-control-theoretical* has reached a very high value asymptotically converging to the optimal, i.e. $\mu^t \gg 0$. Imagine the advertiser realises that they could move the budget to, say, a different marketing channel where they are able to spend the budget. Suddenly, budget becomes the active constraint. Now, $\mu^t$ should go from $\mu^t \gg 0$ to $\mu^t = 0$, which takes a long time. This results in a less efficient behavior from variant-(ii), which ends with a higher cost-per-view. This is shown in Fig 6.

*4.1.8 Practical Scenario 5: Cost is the active constraint and the traffic becomes 2x.* Consider a scenario where cost is the active constraint and that *Cost-control-theoretical* has reached a very high value asymptotically converging to the optimal, i.e. $\mu^t \gg 0$. Assume the traffic on the platform dynamically doubles – this could be because of the traffic variation during the day or due to some special event like a sale. Suddenly, budget becomes the active constraint. As before, $\mu^t$ should go from $\mu^t \gg 0$ to $\mu^t = 0$, which takes a long time. This results in a less efficient behavior from variant-(ii), which ends with a slightly higher cost-per-view. This is shown in Fig 7.

*4.1.9 Practical Scenario 6: Budget is the active constraint and the traffic drops to 0.5x.* Consider a scenario where budget is the active constraint and $\mu^t = \mu^* = 0$ for variants (ii) and (iii). Now consider that the traffic on the platform halves – this could happen due to the traffic variations within a day or a more drastic event like a system failure. Suddenly, cost becomes the active constraint. For *Cost-control-theoretical*, $\mu^* \gg 0$ and the algorithm has to move towards this optimal value. This takes a long time, which leads to a higher violation of the cost-per-view constraint as compared to other variants. This can be observed in Fig 8.

*4.1.10 Summary:* We consider six different practical scenarios. Scenarios 1 and 2 are the most frequent ones. Scenarios 3 and 4, although less frequent, might be critical for a single advertiser who cares about costs (ultimately, return on Ad spend) and is optimizing for it. Scenarios 5 and 6 demonstrate the dynamic nature of the environment and they may be important for designing the algorithm. Moreover, the simulation, here, is simplistic (e.g. all views are equally valuable, view-rate increases smoothly with increasing bid, etc.). In the real-world, the actual impact from different scenarios may be different. This is presented in the next section.

## 4.2 Large-scale Evaluation on Real-world Data

The goal of the large-scale evaluation is to observe to what extent the target costs will be achieved on real-world campaign data and real-world Ad opportunities at scale. Additionally, we also want to observe the utility (cost-adjusted value) generated to the advertisers.

*4.2.1 Set-up:* For large-scale evaluation, we choose a click objective (engagement campaigns) as this differentiates the valuation of views based on their probability to convert to a click. We maximise utility, i.e. $\alpha = 1$. We consider $O(10^7)$ Ad opportunities and $O(10^3)$ Ad campaigns. Each campaign participates in only a fraction of auctions due to targeting. The ads compete with organic content for user views (with organic content effectively acting as floors). For a given page, all variants receive the same number of views (regardless of the ranking of the content). If a page has $K$ views, then they go to the top-K contents on the page. The clicks for each viewed content is sampled from the predicted click-through-rate. This assumes that the user behavior changes only after viewing the content, not before. The auction pricing is second-price. The payment event is view and the advertisers specify a target cost-per-view. The dual variables are updated every 1 minute with learning rate as recommended in [9].

*4.2.2 Variants:* We test six variants: (i) *Max-cap*, (ii) *Cost-control-theoretical*, (iii)-(vi) *Cost-control-practical* with $\beta = 0.8, 0.5, 0.2, 0.0$.

*4.2.3 Metrics:* Let $M$ be the number of campaigns and let $C_{\text{view}}^j$ and $c_{\text{view}}^j$ be the target and actual cost-per-view of the $j^{\text{th}}$ campaign. Then, we measure the number of campaigns for which the cost-per-view constraint is violated by more than 5% of the target cost. That is,

$$M_{\text{violations}} = \frac{1}{M} \sum_{j=1}^{M} \mathbb{1}\left(\frac{c_{\text{view}}^j - C_{\text{view}}^j}{C_{\text{view}}^j} > 0.05\right) \quad (5)$$

Additionally, let $u^j$ be the utility accrued by the $j^{\text{th}}$ campaign. We define *Max-cap* as the baseline variant and measure the relative uplift in the utility. That is,

$$U_{\text{advertiser}} = \frac{1}{M} \sum_{j=1}^{M} \frac{(u_{\text{variant}}^j - u_{\text{baseline}}^j)}{u_{\text{baseline}}^j} \quad (6)$$

We find that the number of views for some campaigns can be very low and the number of clicks can be even lower (usually, 2 orders of magnitude smaller). This results in high variance in the above metrics that are based on cost-per-view and advertiser value (i.e. clicks). Therefore, we keep only those campaigns with more than 300 views and at least, 1 click in the baseline, still leaving $O(10^3)$ Ad campaigns, and we report the metrics on those campaigns.

*4.2.4 Results:* We show the results in Tab 1. From the table, we can see that *Max-cap* has zero violations on the target cost-per-view,

Table 1: Large-scale evaluation results

|  | $M_{violations}$ | $U_{advertiser}$ |
|---|---|---|
| *Max-cap* | 0% | - |
| *Cost-control-theoretical* | 8.15% | +22.09% |
| *Cost-control-practical, $\beta = 0.8$* | 4.12% | +22.60% |
| *Cost-control-practical, $\beta = 0.5$* | 5.13% | +25.84% |
| *Cost-control-practical, $\beta = 0.2$* | 5.21% | +17.49% |
| *Cost-control-practical, $\beta = 0.0$* | 5.46% | +11.68% |

but generates the lowest utility to the advertisers. *Cost-control-theoretical* violates the target cost-per-view on 8.15% of campaigns, but generates +22.09% uplift in utility compared to *Max-cap*. *Cost-control-practical* with $\beta = 0.8, 0.5, 0.2, 0.0$ violate the target cost-per-view on only 4.12%, 5.13%, 5.21%, 5.46% of campaigns respectively while generating +22.60%, +25.84%, +17.49%, +11.68% uplifts in utility over *Max-cap* respectively. Moreover, compared to *Cost-control-theoretical*, *Cost-control-practical* reduces the cost violations by 50% while generating the same utility, with $\beta = 0.8$ making the best trade-off. Based on the results, we also find that *Cost-control-practical* provides a more reliable cost lever to advertisers for steering their campaigns towards the desired volume and efficiency.

## 5 CONCLUDING REMARKS

*Summary:* We considered a scenario where advertisers run a marketing campaign to achieve a certain objective under budget and cost-per-outcome constraints. We showed the ineffectiveness of the theoretically derived bidding formula in meeting the cost constraints and proposed an adjusted bidding formula with a new $\beta$ parameter. Through simulations and large-scale evaluations, we demonstrated that the proposed bidding formula reduces cost violations by 50%.

*How to select $\beta$:* The optimal value of $\beta$ depends on how close the prices are to the bids, which is a property of the market itself. If the market competition is high, then prices (in second-price auctions) closely follow the winning bids. In this case, we need a lower $\beta$ (say, 0.5) to push the bids and correspondingly, the prices down to achieve an effective cost control. On the other hand, if the market competition is low, then prices are much lower than the winning bids. In this case, a $\beta$ slightly lower than 1 (say, 0.9) is already enough to push the prices down and control costs. A $\beta$ close to 0 is not recommended as it negatively impacts the utility generated to the advertiser (as shown in Tab 1). Ultimately, $\beta$ should be viewed as a hyperparameter that is tuned for the market.

*Future work:* While we proposed an alternative bidding formula and proved its effectiveness in an empirical manner, it still lacks a theoretical justification. Moreover, it is also not clear if this modification is the most effective one. Future work will be to explore and compare with other modifications and possibly arrive at a theoretical justification.

## 6 AUTHOR CONTRIBUTIONS

Anoop leads the bidding algorithm design and contributed to setting up the problem, identifying the improvement, running simulations on the synthetic data and producing the results for large-scale evaluation. Rui contributed to identifying the improvement, running simulations on the synthetic data and the engineering implementation. Rinchin contributed to the engineering implementation.

## ACKNOWLEDGMENTS

Thanks to Lorenz Knies for building the tool that enabled the large-scale evaluation. Thanks to Grigorii Fadeev, Dmitry Volodin, Benjamin Tanz, Mustafa Khandwawala and Amir Davari for the valuable discussions. Thanks to the leadership team for sponsoring the work – Manuel Vanzetti, Evgeny Belov, Nicolas Guénon, Paul Gorman, Dinesh Deva. Thanks to Andrey Kashin and Tigran Bagramyan for keeping us true to the customer requirements. Thanks to the reviewers for their valuable feedback.